\documentstyle[12pt,psfig]{article}

\begin{document}

\author{D.V.Dmitriev $^1$ , V.Ya.Krivnov $^2$ and A.A.Ovchinnikov $^{1,2}$ \\
\and $^1$ Max-Planck-Institut fur Physik Komplexer Systeme,
\and Nothnitzer Str. 38, 01187 Dresden, Germany \\
\and $^2$ Joint Institute of Chemical Physics of RAS,
\and Kosygin str.4, 117977, Moscow, Russia}
\title{Exact ground state for 1D electronic models }
\date{}
\maketitle

\begin{abstract}

We have found the exact ground state for two electronic models on a 
linear chain. The first model describes half-filling electron system 
in the ferromagnet--antiferromagnet transition point. In the singlet 
ground state the spin correlators show giant spiral magnetic ordering 
with the period of a spiral equals to the system size. The second 
electronic model describes the point where the ground state has giant 
spiral off-diagonal long-range order and, therefore, superconducting.
We suggest the formation of the ground state with giant spiral order 
(ferromagnetic or off-diagonal) as a probable scenario of the subsequent 
destruction of the ferromagnetism and the superconductivity.

\end{abstract}

\section{Introduction}

In recent years there has been a growing interest in studying systems 
of strongly correlated electrons in relation with high-$T_c$ 
superconductivity. Because of the difficulty in dealing with the many-body 
problems the exact results are rare. It is well-known that some 1D 
electron models can be exactly solved by the Bethe ansatz. However, a lot 
of 1D quantum systems do not obey the Yang-Baxter equation, and thus are 
non-integrable. Another approach leading to exact results consists in 
construction of the exact ground state wave function for some quantum systems. 
Recently considerable progress in this problem has been achieved by 
using the so-called matrix-product (MP) form of the ground state wave function. 
It allowed to find exact ground state for various 1D spin models 
\cite{fannes}-\cite{mik}. Its origin can be traced back to the $S=1$ spin 
chain model \cite{AKLT}. For higher dimensional spin and electronic systems, 
there are also some methods for the construction of the exact ground state 
wave function \cite{JETP}-\cite{Boer}.

There is a class of the 1D quantum spin models describing the ferromagnet
-- antiferromagnet transition point, for which the exact ground state wave 
function has been found in \cite{Japan,pr}. The singlet ground state 
wave function at this point has a special recurrent form, and for special 
values of model parameters it can be reduced to the MP form or to the RVB form
\cite{Japan}. Spin correlations in the singlet ground state show giant spiral 
magnetic structure with the 
period of the spiral equals to the system size. On the antiferromagnet side of 
this point the ground state can be either gapless with the algebraic decay of 
spin correlations \cite{KO} or gapped with the exponential decay of correlations 
\cite{EPJ}. So, this model describes the boundary between the ferromagnetic 
phase and the singlet phase without long-range order.

In this paper we present the singlet ground state wave function of this spin 
model in another form, which can be easily generalized for the 1D electronic
models. Then we consider two 1D electronic models. The first model describes 
half-filling electron system at the point where the singlet and ferromagnetic 
states are degenerate. The exact calculation of the correlation functions in 
the singlet ground state shows the same giant spiral magnetic ordering as for 
the original spin model, while all other correlations vanish in the thermodynamic 
limit. The second electronic model describes the boundary on the phase diagram 
between the superconducting phase with off-diagonal long-range order (ODLRO)
and the non-superconducting phase. The correlation functions in the ground 
state of this model show the giant spiral off-diagonal long-range order. We 
presume that in one-dimensional systems the destruction of the long-range
order (ferromagnetic or off-diagonal) can be followed by the appearance of the 
ground state with the giant spiral order. 

We generalize this form of the wave function for the electronic ladder 
model. This model possesses both the giant spiral spin order and the giant
spiral ODLRO in the ground state. Therefore, this electron ladder model
describes the boundary points on the phase diagram between four different
phases: two singlet phases with and without ODLRO, and two ferromagnetic
phases with and without ODLRO. For some special cases the ground state wave 
function can be reduced to the usual MP form.

The paper is organized as follows. In Section 2 we construct the exact singlet 
ground state for the quantum spin model. In Section 3 two electronic models
with exact ground states are considered, and the correlation functions are
exactly calculated. Section 4 gives a brief summary. In Appendix a technique
for the calculation of correlators is developed.

\section{Quantum spin model}

First, we consider the $s=\frac 12$ spin chain model with nearest- and 
next-nearest neighbor interactions given by the Hamiltonian
\begin{equation}
H=-\sum_{i=1}^{N}({\bf S}_{i}\cdot {\bf S}_{i+1}-\frac{1}{4})+\frac{1}{4}
\sum_{i=1}^{N}({\bf S}_{i}\cdot {\bf S}_{i+2}-\frac{1}{4})  \label{spinh}
\end{equation}
with periodic boundary conditions and even $N$.

This model describes the ferromagnet--antiferromagnet transition point
where the ferromagnetic and singlet states are degenerate.
The Hamiltonian (\ref{spinh}) has been considered in \cite{Japan,pr},
where the singlet ground state wave function was constructed in two 
different forms. In this paper we represent another form of this singlet 
function, which allows us to generalize this function for the electronic 
model and to develop a technique to calculate correlators. 

The singlet ground state wave function for the Hamiltonian (\ref{spinh})
can be written as follows:
\begin{equation}
\Psi _{0} = P_{0}\Psi , \qquad
\Psi = \langle 0_b|\, g_1\otimes g_2\otimes ...\otimes g_N \, |0_b \rangle  
\label{swf}
\end{equation}
where
\begin{equation}
g_i = b^{+}\,\left|\uparrow\right\rangle _i 
+ b \, \left|\downarrow\right\rangle _i
\end{equation}
Here we introduced one auxiliary Bose-particle $b^{+},\;b$ (the Bose operators
$b^{+},\;b$ do not act
on spin states $\left|\uparrow\right\rangle _i$ and $\left|\downarrow\right
\rangle _i$) and the Bose vacuum $\left|0_b\right\rangle$. Therefore, the
direct product $g_1\otimes g_2\otimes ...\otimes g_N$ is the superposition of
all possible spin configurations multiplied on the corresponding Bose operators, like 
$b^{+}b\,b\,b^{+}\, ...\,\left|\uparrow\downarrow\downarrow\uparrow ...\right\rangle$.
$P_0$ is a projector onto the singlet state.
This operator can be written as \cite{P0}
\begin{equation}
P_0=\frac 1{8\pi ^2}\int_0^{2\pi }d\alpha \int_0^{2\pi }d\beta \int_0^\pi
\sin \gamma d\gamma ~e^{i\alpha S^z}e^{i\gamma S^x}e^{i\beta S^z},
\label{P0}
\end{equation}
where $S^{x(z)}$ are components of the total spin operator.

This form of wave function resembles the MP form but with infinity matrix,
which is represented by Bose operators. Therefore, we have to pick out the
$\langle 0_b|...|0_b \rangle $ element of the matrix product instead of 
usual $Trace$ in the MP formalism \cite{fannes}-\cite{mik}, because 
$Trace$ is undefined in this case. 
The function $\Psi $ contains components with all possible values of 
spin $S$ ($0\leq S\leq N/2$) and, in fact, a fraction of the singlet is
exponentially small at large $N$. This component is filtered out by the
operator $P_0$.

In order to show that $\Psi_0 $ is the ground state wave function for 
the Hamiltonian (\ref{spinh}), let us represent the Hamiltonian (\ref{spinh})
as a sum of Hamiltonians $h_i$ of cells containing three sites
\begin{equation}
H=\sum_{i=1}^N h_i ,  \label{3}
\end{equation}
where
\[
h_i = -\frac 12({\bf S}_{i}{\bf S}_{i+1}-\frac 14)-\frac 12({\bf S
}_{i+1}{\bf S}_{i+2}-\frac 14)+\frac 14({\bf S}_{i}{\bf S}_{i+2}
-\frac 14)
\]

One can easily check that each cell Hamiltonian $h_i$ for $i=1,..N-2$
gives zero when acting on $g_i\otimes g_{i+1}\otimes g_{i+2}$.

Since each $h_i$ is non-negatively defined operator, then $\Psi $ is 
the exact ground state wave function of an open chain:
\[
H_{\rm open}=\sum_{i=1}^{N-2} h_i
\] 

As it was mentioned above, the function $\Psi $ contains components 
with all possible values of total spin $S$, and, therefore, the ground 
state of the open chain is multiply degenerate. But it can be proven
(as it was done in \cite{pr}),
that for the cyclic chain (\ref{spinh}) only singlet and ferromagnetic 
components of $\Psi $ have zero energy. Therefore, for the cyclic 
chain (\ref{spinh}) $\Psi _0$ is the singlet ground state wave 
function degenerate with the ferromagnetic state.

The exact calculation of the norm and spin correlation function $\langle {\bf
S}_i {\bf S}_{i+l}\rangle$ (see Appendix) 
in the singlet ground state (\ref{swf}) results in the following expressions:
\begin{eqnarray}
\langle \Psi _0| \Psi _0\rangle 
=\left. \frac {d^N}{d\xi ^N} \left( \frac {1}{\cos^2(\frac{\xi }{2})} 
\right)  \right|_{\xi =0}
\label{snorma1t}
\end{eqnarray}
\begin{equation}
\langle \Psi _0 |{\bf S}_i {\bf S}_{i+l}| \Psi _0\rangle 
= \frac {\partial ^l}{\partial \xi ^l} \frac {\partial ^{N-l-2}}
{\partial \zeta ^{N-l-2}} \left. \left( -\frac 38 \frac 
{\cos (\xi -\zeta )}{\cos^4(\frac{\xi +\zeta }{2})} \right)  
\right|_{\xi =\zeta =0}
\label{spincorrt}
\end{equation}

It can be shown that in the thermodynamic limit equations (\ref{snorma1t}) and
(\ref{spincorrt}) result in
\begin{equation}
\left\langle {\bf S}_{i}{\bf S}_{i+l}\right\rangle =
\frac{1}{4}\cos \left( \frac{2\pi l}{N}\right)   
\label{spincos} 
\end{equation}
So, we reproduce the result obtained in \cite{Japan,pr} that
in the thermodynamic limit a giant spiral spin structure is realized, 
with the period of spiral equal to the system size.

\section{Electronic models}

Now we will construct electronic models by generalization of the wave 
function (\ref{swf}):
\begin{equation}
\Psi_0 = P_{S=0}\; Tr_f \; \langle 0_b|\, 
g_1\otimes g_2\otimes ...\otimes g_N \, |0_b \rangle 
\label{e1wf}
\end{equation}
where
\begin{equation}
g_i = b^{+}\left|\uparrow \right \rangle _i
+ b\left|\downarrow \right \rangle _i
+c\:(f^{+} \left| 2 \right \rangle _i +f \left| 0 \right \rangle _i )
\label{e1g}
\end{equation}
with $b^{+},\,b$ and $f^{+},\,f$ are the Bose and the Fermi operators 
respectively, $\left|0_b\right\rangle $ is the Bose vacuum, and $c$ is a
parameter of the model. Here we also denote an empty site by $\left| 0 
\right \rangle$, a site occupied by one electron by $\left|\uparrow \right 
\rangle$ and $\left|\downarrow \right \rangle$ and a doubly occupied site 
by $\left| 2 \right \rangle$. So, the  product $g_1\otimes ...g_N$ is the 
operator in the Bose $b^{+},\,b$ and the Fermi $f^{+},\,f$ spaces. 
We pick out the $\langle 0_b|...|0_b \rangle$ element in the Bose space, 
which can be written as:
\begin{equation}
\langle 0_b|\,g_1\otimes g_2\otimes ...\otimes g_N \, |0_b \rangle
= \phi_0 + (f^+f-ff^+)\:\phi_1
\label{phi}
\end{equation}
and then we take the $Trace$ over the Fermi operators:
\[
Tr_f \langle 0_b|\,g_1\otimes g_2\otimes ...\otimes g_N \, |0_b \rangle 
= 2\phi_0 
\]
The projector $P_{S=0}$ filters out the singlet component from the
function $\phi_0$.

Thus, we have a singlet wave function $\Psi_0$ describing the state 
with one electron per site.

In order to find the Hamiltonian for which the wave function (\ref{e1wf}) 
is the exact ground state wave function, let us consider what 
states are present on the two nearest sites in the wave function (\ref{e1wf}).
One can easily check that there are only 9 states from the total 16 states
in the  product $g_i\otimes g_{i+1}$. They are
\begin{eqnarray}
&&\left|\uparrow \uparrow \right \rangle ,\qquad
\left|\downarrow \downarrow  \right \rangle ,\qquad
\left|\uparrow \downarrow + \downarrow \uparrow  \right \rangle ,
\qquad \left|20-02 \right \rangle ,\qquad
\left|\uparrow \downarrow - \downarrow \uparrow  \right \rangle
- c^2 \: \left|20+02 \right \rangle ,
\nonumber \\
&&\left|\uparrow 0 - 0\uparrow \right \rangle ,\qquad
\left|\uparrow 2 - 2\uparrow \right \rangle ,\qquad
\left|\downarrow 0 - 0\uparrow \right \rangle ,\qquad
\left|\downarrow 2 - 2\downarrow \right \rangle 
\label{states1}
\end{eqnarray}

The elementary Hamiltonian $h_{i,i+1}$ for which all these states
are the exact ground states can be written as the sum of the projectors 
onto the 7 missing states $\left|\varphi_k \right \rangle$ 
with arbitrary positive coefficients $\lambda _k$
\[
h_{i,i+1}=\sum\limits_{k=1}^{7}\lambda _{k}  
\left|\varphi_k \right \rangle  \left \langle \varphi_k \right|
\]

For $c>1$ the most simple form of this Hamiltonian is:
\begin{eqnarray}
H &=& \sum_{i=1}^{N} h_{i,i+1} \label{e1h} \\
h_{1,2} &=& 1-4\:{\bf S}_{1}\cdot {\bf S}_{2} +4\,
(1-\frac{3}{c^4})\eta_1^z\eta_2^z
+\frac{4}{c^4}{\bf \eta}_{1}\cdot {\bf \eta}_{2}
\nonumber \\
&+& \sum_{\sigma}(c_{1,\sigma}^{+}c_{2,\sigma}+c_{2,\sigma}^{+}c_{1,\sigma})
(1-n_{1,-\sigma}-n_{2,-\sigma})
\nonumber \\
&+& \frac{2}{c^2}\sum_{\sigma}(c_{1,\sigma}^{+}c_{2,\sigma}+
c_{2,\sigma}^{+}c_{1,\sigma})(n_{1,-\sigma}-n_{2,-\sigma})^2
\nonumber
\end{eqnarray}
Here $c_{i,\sigma}^{+}$, $c_{i,\sigma}$ are the Fermi operators,
$n_{i,\sigma}=c_{i,\sigma}^{+}c_{i,\sigma}$, and the $SU(2)$ spin 
operators are given by $S_i^{+}=c_{i,\uparrow}^{+}\: c_{i,\downarrow}$,
$S_i^{-}=c_{i,\downarrow}^{+}\: c_{i,\uparrow}$ and
$S_i^{z}=\frac 12 (n_{i,\uparrow }-n_{i,\downarrow })$.
We also use here ${\bf \eta}$ operators:
\[
\eta_i^{+}=c_{i,\downarrow }^{+}\:c_{i,\uparrow }^{+}, \qquad
\eta_i^{-}=c_{i,\uparrow }\:c_{i,\downarrow }, \qquad
\eta_i^{z}=\frac {1-n_{i,\uparrow }-n_{i,\downarrow }}{2} , 
\]
which form another $SU(2)$ algebra \cite{yang1,eta}, and 
${\bf \eta}_{1}\cdot {\bf \eta}_{2}$ is a scalar product of pseudo-spins
${\bf \eta}_{1}$ and ${\bf \eta}_{2}$.

The Hamiltonian (\ref{e1h}) does not conserve the total number of empty and
doubly occupied sites because of the last term in the elementary Hamiltonians 
$h_{i,i+1}$ , in contrast to the models considered in \cite{eta}.

Each elementary Hamiltonian $h_{i,i+1}$ $(i=1,..N-1)$ acting on functions 
$\phi_0$, $\phi_1$ gives zero, since all the states (\ref{states1})
are the eigenstates of $h_{i,i+1}$ with zero energy, while the energies of all 
other states at $c>1$ are positive. Therefore, the functions $\phi_0$ 
and $\phi_1$ are the ground state wave functions of the open chain:
\begin{equation}
H_{\rm open} = \sum_{i=1}^{N-1} h_{i,i+1}
\label{e1hopen}
\end{equation}

To determine the degeneracy of the model (\ref{e1hopen}), we need to classify
the functions $\phi_n$. Analogously to the spin model (\ref{swf}),
the functions $\phi_0$ and $\phi_1$ contain components with all possible 
values of total spin $S$. Therefore, $\phi_0$ contains multiplets with   
$S=0,...N/2$ and $\phi_1$ contains components with values of the total 
spin $S=0,...N/2-1$ ($\phi_1$ does not contain the ferromagnetic component,
since at least two sites in $\phi_1$ are non-magnetic
$\left|0 \right \rangle$ and $\left|2 \right \rangle$). 
So, $N+1$ multiplets are degenerated for the open chain.

But for the cyclic model (\ref{e1h}) it can be proved that only three
multiplets are the ground states: singlet state (\ref{e1wf}) with the momentum 
$p=0$ (singlet component of $\phi_0$), the trivial ferromagnetic state with 
$p=\pi$, and the state with $S=N/2-1$ and $p=\pi$ (which is the component of 
$\phi_1$ with $S=N/2-1$). The last state with $S=N/2-1$ can be written as:
\[
\Psi_{N/2-1} =\sum_{i<j} (c_{i,\uparrow }^{+}c_{j,\downarrow } +
c_{j,\uparrow }^{+}c_{i,\downarrow })\;\prod_{n=1}^{N} c_{n,\downarrow }^{+}
\left|0 \right \rangle
\]

Thus, the ground states of the electronic model (\ref{e1h}) with one electron per 
site are the singlet state, the ferromagnetic state, and the state with 
$S=S_{\rm max}-1$.

It is interesting to note that the singlet wave function (\ref{e1wf}) 
can be also written in the form:
\[
\Psi _0 =\sum [i,j][k,l][m,n]...\;\prod_{n=1}^{N} c_{n,\downarrow }^{+}
\left|0 \right \rangle
\]
where 
\[
[i,j] = S_i^{+} - S_j^{+} + c^2\:(c_{i,\uparrow }^{+}c_{j,\downarrow }
- c_{j,\uparrow }^{+}c_{i,\downarrow }) ,
\]
and the summation is made for any combination of sites under the condition 
that $i<j,$ $k<l,$ $m<n$ $...$. This form of the wave function is analogous
to the RVB form found in \cite{Japan} for the spin model (\ref{spinh}). 

The norm and the correlators of the electronic model (\ref{e1h}) in the
singlet ground state are calculated in the same way as for the spin model 
(Appendix). Therefore, here we give the final results:
\begin{eqnarray}
\langle \Psi _0| \Psi _0\rangle  &=&
\left. \frac {d^{N}}{d\xi ^{N}} 
\left( 2\frac{1+\cosh(c^2\xi )}{\cos^2(\frac{\xi}{2})}  \right)  \right|_{\xi=0}
\label{e1norma} \\
\langle \Psi _0 |{\bf S}_i {\bf S}_{i+l}| \Psi _0\rangle 
&=& \frac {\partial ^l}{\partial \xi ^l} \frac {\partial ^{N-l-2}}
{\partial \zeta ^{N-l-2}} \left. \left( -\frac 34 \frac 
{\cos (\xi -\zeta ) (1+\cosh (c^2\xi +c^2\zeta ))}
{\cos^4(\frac{\xi +\zeta }{2})} \right)  
\right|_{\xi =\zeta =0}
\label{espincorr} \\
\langle \Psi _0 |c_{i,\sigma}^{+}\:c_{i+l,\sigma}| \Psi _0\rangle 
&=& \frac {\partial ^l}{\partial \xi ^l} \frac {\partial ^{N-l-2}}
{\partial \zeta ^{N-l-2}} \left. \left( -c^2 \frac 
{(\cos(\xi )+cos(\zeta )) (\cosh(c^2\xi )+\cosh(c^2\zeta ))}
{2\cos^4(\frac{\xi +\zeta }{2})} \right)  
\right|_{\xi =\zeta =0}
\label{tcorr} \\
\langle \Psi _0 |\eta_i^{z} \eta_{i+l}^{z}| \Psi _0\rangle 
&=& \frac {\partial ^l}{\partial \xi ^l} \frac {\partial ^{N-l-2}}
{\partial \zeta ^{N-l-2}} \left. \left( 
- \frac {c^4 \cosh(c^2\xi -c^2\zeta )}
{2\cos^2(\frac{\xi +\zeta }{2})} \right)  
\right|_{\xi =\zeta =0}
\label{etazcorr} \\
\langle \Psi _0 |\eta_i^{-} \eta_{i+l}^{+}| \Psi _0\rangle 
&=& \left. \frac {d^{N-2}}{d\xi ^{N-2}} 
\left( \frac{c^4}{\cos^2(\frac{\xi}{2})}  \right)  \right|_{\xi=0}
\label{etacorr}
\end{eqnarray}

As it can be seen from Eq.(\ref{etacorr}), the expectation value 
$\langle \eta_1^{-} \eta_{l+1}^{+} \rangle $, which determines
the off-diagonal long-range order (ODLRO) \cite{yang}, does not depend on 
the distance $l$. But in the thermodynamic limit ODLRO vanishes:
\begin{eqnarray}
\langle \eta_i^z \eta_{i+l}^z \rangle &=& O(\frac {1}{N^2}) ,\qquad
\langle \eta_i^+ \eta_{i+l}^- \rangle = O(\frac {1}{N^2}) ,\nonumber \\
\langle c_{i,\sigma}^{+}\:c_{i+l,\sigma}\rangle &=& O(\frac 1N) ,
\qquad
\left\langle {\bf S}_{i}{\bf S}_{i+l}\right\rangle =
\frac{1}{4}\cos \left( \frac{2\pi l}{N}\right) ,
\label{spincos1} 
\end{eqnarray}
though for finite systems all correlators (\ref{tcorr})-(\ref{etacorr}) are
non zero. 

The second electronic model can be obtained by simply interchanging of the
Bose and the Fermi operators in (\ref{e1g}). So, the wave function of this
model has the form:
\begin{equation}
\Psi _{0}=P_{\eta =0}\; Tr_f \; \langle 0_b|\, 
g_1\otimes g_2\otimes ...\otimes g_N \, |0_b \rangle 
\label{e2wf}
\end{equation}
with
\begin{equation}
g_i = c\:(f^{+}\left|\uparrow\right\rangle _i
+ f\, \left|\downarrow \right \rangle _i)
+b^{+} \left|2\right\rangle _i + b\left|0\right\rangle _i
\label{e2g}
\end{equation}
The projector $P_{\eta =0}$ filters out the state with total
${\bf \eta}=\sum{\bf \eta}_i=0$.
Therefore, the function $\Psi_0$ has $S^z=0$, but it is not an eigenfunction of
$S^2$. Instead, it is an eigenfunction of ${\bf \eta}^2$ with ${\bf \eta}=0$. 

The wave function (\ref{e2wf}) can be also written in the form analogous
to the RVB one:
\[
\Psi _0 =\sum [i,j][k,l][m,n]...\left|0 \right \rangle
\]
where 
\[
[i,j] = \eta_i^{+} - \eta_j^{+} + c^2\:(c_{i,\uparrow }^{+}c_{j,\downarrow }^{+}
+ c_{i,\downarrow }^{+}c_{j,\uparrow }^{+}) ,
\]
and the summation is also done over any combinations of sites under the condition 
that $i<j,$ $k<l,$ $m<n$ $...$.

Considering the product $g_i\otimes g_{i+1}$ one can find that there are
only following 9 states on two nearest sites in the wave function (\ref{e2wf}) 
\begin{eqnarray}
&&\left| 22 \right \rangle ,\qquad
\left| 00  \right \rangle ,\qquad
\left| 20+02  \right \rangle ,\qquad
\left|\uparrow \downarrow - \downarrow \uparrow \right \rangle ,\qquad
\left|20-02 \right \rangle
+ c^2 \:\left|\uparrow \downarrow + \downarrow \uparrow  \right \rangle ,
\nonumber \\
&&\left|\uparrow 0 + 0\uparrow \right \rangle ,\qquad
\left|\uparrow 2 + 2\uparrow \right \rangle ,\qquad
\left|\downarrow 0 + 0\uparrow \right \rangle ,\qquad
\left|\downarrow 2 + 2\downarrow \right \rangle 
\label{states2}
\end{eqnarray}

The Hamiltonian for this model has the form which is similar to the previous
one given by Eq.(\ref{e1h}):
\begin{eqnarray}
H &=& \sum_{i=1}^{N} h_{i,i+1} \label{e2h} \\
h_{1,2} &=& 1-4{\bf \eta}_{1}\cdot {\bf \eta}_{2} +4(1-\frac{3}{c^4})S_1^z S_2^z
+\frac{4}{c^4}{\bf S}_{1}\cdot {\bf S}_{2}
\nonumber \\
&-&\sum_{\sigma}(c_{1,\sigma}^{+}c_{2,\sigma}+c_{2,\sigma}^{+}c_{1,\sigma})
(1-n_{1,-\sigma}-n_{2,-\sigma})
\nonumber \\
&+&\frac{2}{c^2}\sum_{\sigma} \sigma(c_{1,\sigma}^{+}c_{2,\sigma}+
c_{2,\sigma}^{+}c_{1,\sigma})(n_{1,-\sigma}-n_{2,-\sigma})^2
\nonumber
\end{eqnarray}

This Hamiltonian for $c>1$ is also a non-negatively defined operator and 
$\Psi_0$ is the exact ground state wave function with zero energy.
This Hamiltonian commutes with $\eta^2$, but does not commute with $S^2$.
Therefore, the eigenfunctions of the Hamiltonian (\ref{e2h}) can be
described by quantum numbers ${\bf\eta}$ and $\eta^z$. Making up the 
same analysis as for previous model we find that for the cyclic model 
(\ref{e2h}) the states with three different values of $\eta$ have zero 
energy (as it was for model (\ref{e1h})). They are:
one state with $\eta =0$ and momentum $p=\pi$ (\ref{e2wf}), 
all states with $\eta =N/2$ and $p=0$:
\begin{equation}
\Psi_{N/2,\;\eta^z} = (\eta^+)^{N/2-\eta^z} \left|0 \right \rangle
\label{e2wf2}
\end{equation}
and the states with $\eta =N/2-1$ and $p=0$:
\begin{equation}
\Psi_{N/2-1,\;\eta^z} = (\eta^+)^{N/2-1-\eta^z} \,
\sum_{i<j} (c_{i,\uparrow }^{+}c_{j,\downarrow }^{+}
- c_{i,\downarrow }^{+}c_{j,\uparrow }^{+}) \;\left|\,0 \right \rangle
\label{e2wf3}
\end{equation}

Therefore, for the case of one electron per site ($\eta^z=0$) the ground 
state of the model (\ref{e2h}) is three-fold degenerate.

The correlation functions in the ground states (\ref{e2wf2}) and (\ref{e2wf3}) 
obviously coincide with each other in the thermodynamic limit and for half 
filling case ($\eta^z=0$) they are:
\begin{eqnarray}
\langle c_{i,\sigma}^{+}\:c_{i+l,\sigma}\rangle = O(\frac 1N)  \qquad
\langle {\bf S}_{i}{\bf S}_{i+l} \rangle = O(\frac {1}{N^2}) \nonumber \\
\left\langle \eta_i^z \eta_{i+l}^z \right\rangle = O(\frac 1N) \qquad
\left\langle \eta_i^- \eta_{i+l}^+ \right\rangle = \frac{1}{4} + O(\frac 1N) 
\label{etaferro} 
\end{eqnarray}
The existence of ODLRO immediately follows from the form of the wave
functions (\ref{e2wf2}) and (\ref{e2wf3}).

The correlation functions in the ground state (\ref{e2wf}) have similar 
forms as in Eqs.(\ref{espincorr})--(\ref{etacorr}), and in the thermodynamic 
limit they reduce to
\begin{eqnarray}
\langle c_{i,\sigma}^{+}\:c_{i+l,\sigma}\rangle = O(\frac {1}{N}),
\qquad \langle {\bf S}_{i}{\bf S}_{i+l} \rangle = O(\frac {1}{N^2}),
\nonumber \\
\left\langle \eta_i^- \eta_{i+l}^+ \right\rangle =
2 \left\langle \eta_i^z \eta_{i+l}^z \right\rangle =
\frac{1}{6}\cos \left( \frac{2\pi l}{N}\right) 
\label{etacos} 
\end{eqnarray}

The giant spiral ordering in the last equation implies the existence 
of ODLRO and, therefore, the superconductivity \cite{yang} in the ground state 
(\ref{e2wf}). We note that though all three ground states of the model 
(\ref{e2h}) are superconducting, the properties of these wave functions are 
essentially different. Let us consider density -- density correlator 
$\left\langle n_i n_{i+l} \right\rangle$. For the wave functions (\ref{e2wf2}) 
and (\ref{e2wf3}) in the thermodynamic limit this correlator decouples:
$\left\langle n_i n_{i+l} \right\rangle = 
\left\langle n_i \right\rangle \left\langle n_{i+l} \right\rangle =1$.
But for the wave function (\ref{e2wf}) it equals to $\left\langle n_i n_{i+l} 
\right\rangle = 1+\frac{1}{3}\cos \left( \frac{2\pi l}{N}\right)$.

It is interesting to note that another model having the ground state wave 
function (\ref{e2wf}) with $c=0$ and the same spiral ODLRO (\ref{etacos}) 
can be obtained from the model (\ref{spinh}) by simple replacing of operators 
$S$ by $\eta$:
\begin{equation}
H=-\sum_{i=1}^{N} {\bf\eta}_{i}\cdot {\bf\eta}_{i+1} + \frac{1}{4}
\sum_{i=1}^{N} {\bf\eta}_{i}\cdot {\bf\eta}_{i+2} 
\label{e3h}
\end{equation}

The direct analogy of this model to the spin model (\ref{spinh}) results 
in the conclusion that the model (\ref{e3h}) describes the boundary point 
on the phase diagram between superconducting and non -- superconducting
phases where the off-diagonal long-range order is destroyed. We suppose that 
the model (\ref{e2h}) also describes such a point. Thus, the wave functions 
(\ref{e1wf}) and (\ref{e2wf}) are the ground states for the 1D electronic 
systems in the boundary points between the phases with and without long-range 
order (ferromagnetic for (\ref{e1h}) and off-diagonal for (\ref{e2h})).
And we suggest the formation of the ground state with the long-range spiral 
order like (\ref{spincos1}) and (\ref{etacos}) as a probable scenario 
of the subsequent destruction of the ferromagnetism and the superconductivity.

\begin{figure}[t]
\unitlength1cm
\begin{picture}(11,6)
\centerline{\psfig{file=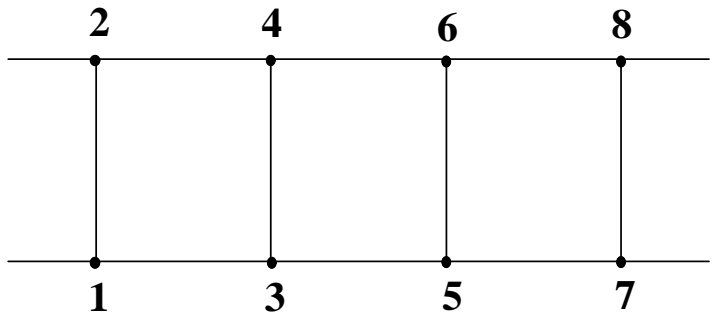,angle=-90,width=10cm}}
\end{picture}
\vspace{-2cm}
\caption{ \label{sss}
The two-leg ladder
}
\end{figure}

The proposed form of wave function can be further generalized for the electronic
ladder model (Fig.1). The wave function for the cyclic ladder model 
containing $2N$ sites has the form:
\begin{equation}
\Psi_0 = P_{S=\eta =0} \; \langle 0|\, 
g_1\otimes g_2\otimes ...\otimes g_N \, |0 \rangle 
\label{lwf}
\end{equation}
where each $g_i$ corresponds to $i$-th rung of the ladder:
\begin{eqnarray}
g_i &=& c_1\;\left( a^+ (2x-a^+a) \left|\uparrow\uparrow \right \rangle _i
- a \left|\downarrow\downarrow \right \rangle _i
+(a^+a-x)\left|\uparrow\downarrow + \downarrow\uparrow\right \rangle _i
\right)
+c_2 \left|\uparrow\downarrow - \downarrow\uparrow\right \rangle _i 
\nonumber\\
&+& c_3\;\left( b^+ (2y-b^+b) \left|22 \right \rangle _i
- b \left|00 \right \rangle _i 
+(b^+b-y)\left| 20+02 \right \rangle _i \right)
+c_4 \left| 20-02 \right \rangle _i
\label{lg}
\end{eqnarray}
with $a^{+},\,a$ and $b^{+},\,b$ are the Bose operators, $\left|0\right
\rangle $ in (\ref{lwf}) is the Bose vacuum of $a^{+}$ and $b^{+}$ 
particles, and $c_i$, $x$, $y$ are the parameters of the model.
$P_{\eta =S=0}$ is the projector onto the state with $S=\eta =0$.

Let us first consider the case $c_3=c_4=0$. In this case the wave function
(\ref{lwf}) describes the spin ladder model depending on the two parameters
$c_2/c_1$ and $x$. It can be shown that this model coincides with that considered
in \cite{EPJ}. It has the singlet ground state (\ref{lwf}) degenerate
with ferromagnetic state. The spin correlators in the singlet ground state
show double-spiral ordering with small shift angle
$\triangle \varphi = \frac{2\pi }{N}\frac{2c_2}{c_1}$
between the two giant spirals formed on two legs of the ladder:
\begin{eqnarray}
\left\langle {\bf S}_{n}{\bf S}_{n+2l}\right\rangle &=&
\frac{1}{4}\cos \left( \frac{2\pi l}{N}\right) \nonumber \\
\left\langle {\bf S}_{n}{\bf S}_{n+2l+1}\right\rangle &=&
\frac{1}{4}\cos \left( \frac{2\pi l}{N}+(-1)^n\triangle \varphi \right)
\label{slcorr}
\end{eqnarray}

For the cases of integer or half-integer $x=j$, which correspond to the 
special cases of the model \cite{EPJ}, one can easily recognize in 
Eq.(\ref{lg}) the Maleev's boson representation of spin $S=j$ operators:
\[
S^+=a^+(2j-a^+a),\qquad S^-=a, \qquad S^z=a^+a-j 
\]
Therefore, in these special cases the infinite matrices formed by the Bose 
operators $a^{+},\,a$ can be broken off to the size $n=2j+1$ and the wave 
function (\ref{lwf}) is reduced to the usual MP form. The spin correlators 
in the special cases have the exponential decay.

Now let us come back to the general case of the electronic ladder model 
(\ref{lwf}). In order to find the Hamiltonian for which (\ref{lwf}) is the 
exact ground state wave function, one should consider what states are present 
in the $\Psi _0$ on the two nearest rungs of the ladder. There are only 26 states 
from the total 256 states in the  product $g_i\otimes g_{i+1}$. Therefore, the 
Hamiltonian of the ladder model can be written as the sum of the projectors 
onto the 230 missing states $\left|\varphi_k \right \rangle$ with arbitrary 
positive coefficients $\lambda _k$:
\begin{equation}
H = \sum_{i=1}^{N} h_{i,i+1},   \qquad
h_{i,i+1}=\sum\limits_{k=1}^{230}     \lambda _{k}  
\left|\varphi_k \right \rangle  \left \langle \varphi_k \right|
\label{lh}
\end{equation}

Unfortunately, we can not give the explicit form like (\ref{e1h}) and
(\ref{e2h}) for this Hamiltonian, because it has a very cumbersome form.
But we are able to determine some properties of the Hamiltonian (\ref{lh}). 
This Hamiltonian commutes with both $S^2$ and $\eta^2$. It has multiply
degenerated ground state: the state with $S=\eta =0$ (\ref{lwf}) and all the 
states with $S+\eta =N$ have zero energy. Hence the electronic ladder model 
(\ref{lh}) describes the boundary point between the phases with and without 
ferromagnetic and off-diagonal long-range order. The correlation functions
in the ground state (\ref{lwf}) can be calculated with the use of the technique 
developed in Appendix. For the case $c_1>c_3$ there is the same double-spiral 
spin ordering (\ref{slcorr}) as for the spin ladder model, while all other
correlations are exponentially small. For the case $c_1<c_3$ the double-spiral 
ODLRO is realized. In the most interesting symmetric case $c_1=c_3$, $c_2=c_4$, 
$x=y$ the system possesses both the giant spiral spin order and the giant spiral 
ODLRO:
\begin{eqnarray}
\left\langle {\bf S}_{n}{\bf S}_{n+2l}\right\rangle &=&
\left\langle {\bf \eta}_{n}{\bf \eta}_{n+2l}\right\rangle =
\frac 18 \cos \left( \frac{2\pi l}{N}\right)  \nonumber \\
\left\langle {\bf S}_{n}{\bf S}_{n+2l+1}\right\rangle &=&
\left\langle {\bf \eta}_{n}{\bf \eta}_{n+2l+1}\right\rangle =
\frac 18 \cos\left(\frac{2\pi l}{N}+(-1)^n \triangle \varphi \right)
\label{lcorr}
\end{eqnarray}
where $\triangle \varphi = \frac{2\pi }{N}\frac{2c_2}{c_1}$.

Therefore, in this case the wave function (\ref{lwf}) describes the boundary 
points on the phase diagram between the four different phases: the singlet phases 
with and without ODLRO, and the ferromagnetic phases with and without ODLRO.
In the special cases when $x$ ($y$) is integer or half-integer the spin
(off-diagonal) correlations exponentially decay and the wave function in the
corresponding Bose space can be represented in the MP form with finite
matrices of the size $n=2x+1$ or $n=2y+1$. When both $x$ and $y$ are integers 
or half-integers the wave function (\ref{lwf}) can be written in the usual MP 
form with the size of matrices $n=(2x+1)(2y+1)$.

\section{Summary}

We have found another form of the singlet ground state wave function for the
quantum spin model considered previously in \cite{Japan,pr}.  The special 
technique was developed for the exact calculation of the norm and the correlation 
functions. This form of the wave function allowed us to generalize it for two 
1D electronic models. 

The first model describes the half-filling electronic system in the 
ferromagnet -- antiferromagnet transition point when the singlet and the
ferromagnetic states are degenerate. In the singlet ground state the spin 
correlators show the giant spiral magnetic ordering with the period of the 
spiral equals to the system size, while all other correlations vanish in 
the thermodynamic limit. 

The second electronic model in the half filling case has three-fold degenerate 
ground state. All ground states have off-diagonal long-range order and, 
therefore, are superconducting. The calculation of the correlation functions 
shows that one of the ground states has a giant spiral ODLRO.
 
The comparison of these electronic models with original spin model \cite{KO,EPJ} 
leads us to the conclusion that these two electronic models describe the boundary 
points on the phase diagram between the phases with and without long-range 
order (ferromagnetic for the first and off-diagonal for the second model). 
Therefore, we presume that if the Hamiltonian of the 1D quantum system 
commutes with operators forming the $SU(2)$ algebra (it can be the spin $S$ or 
the pseudo-spin $\eta$ operators), then the appearance of the ground state with the 
giant spiral order predicts the following destruction of the ferromagnetism or 
the superconductivity. 

We have briefly considered the generalization of the proposed form of the wave
function for the electronic ladder model. The general case of this model has much 
more rich phase diagram than two first models. In some particular cases 
this model describes the boundary points on the phase diagram between four 
different phases: with and without ferromagnetic and off-diagonal long-range
order. There are also some special cases of the electronic ladder model when 
the ground state wave function is reduced to the usual MP form.
Besides, the proposed form of the wave function can be also generalized for the
2D case and different types of lattices.

\section{Acknowledgements}
Authors are grateful to Max-Planck-Institut fur Physik Komplexer 
Systeme for kind hospitality. This work was supported by RFFR.

\setcounter{equation}{0}
\renewcommand{\theequation}{A.\arabic{equation}}
\section*{Appendix}

Let us calculate the norm of the wave function $\Psi _{0}$ (\ref{swf})
\begin{equation}
\langle \Psi _0| \Psi _0\rangle =\langle \Psi |\: P_0\:| \Psi \rangle
\label{snorm0}
\end{equation}
Since the function $\Psi $ has $S^z=0$, then the projector $P_0$ in the 
Eq.(\ref{P0}) takes the form \cite{pr}
\begin{equation}
P_0=\frac 12\int_0^\pi \sin \gamma d\gamma \,  
e^{izS^{-}} e^{iz^{\prime }S^{+}}
\end{equation}
where $z=\tan \frac{\gamma }{2},\; z^{\prime }=\sin \frac{\gamma }{2} 
\cos \frac{\gamma }{2}$ and $S^{+(-)}$ are the operators of the total spin.

Therefore, the norm takes the form:
\[
\langle \Psi _0| \Psi _0\rangle = \frac 12 \int_0^\pi \sin \gamma d\gamma \,
\langle 0_a , 0_b | \prod_{i=1}^{N} (g_{i}^{+} \, e^{izS_i^{-}} e^{iz^{\prime
}S_i^{+}} g_{i}) |0_a , 0_b  \rangle 
\]
where
\begin{eqnarray*}
g_{i}^{+} \, e^{izS_i^{-}} e^{iz^{\prime }S_i^{+}} g_{i}
&=& (a^{+} \left\langle\uparrow _i\right| + a \left\langle\downarrow _i\right|
\,) \, e^{izS_i^{-}} e^{iz^{\prime }S_i^{+}}
(b^{+}\left|\uparrow\right\rangle _i +b\left|\downarrow\right\rangle _i \,)  \\
&=& a^{+}b^{+}+(1-z^{\prime }z)ab+izab^{+}+iz^{\prime }a^{+}b
\end{eqnarray*}
and $a^{+},\;a$ are the Bose operators.

So, the norm can be rewritten as follows 
\begin{equation}
\langle \Psi _0 | \Psi _0\rangle =\frac 12 \int_0^\pi \sin \gamma d\gamma \,
\langle 0| G^N |0\rangle   \label{snorma}
\end{equation}
with $\left|0\right\rangle=\left|0_a ,0_b\right\rangle$ is the Bose vacuum of
$a^{+}$ and $b^{+}$ particles and
\[
G=u(a^{+}b^{+}+ab)+iv(ab^{+}+a^{+}b)
\]
where $u=\cos \frac{\gamma}{2}, \; v=\sin \frac{\gamma}{2}$.

Let us introduce the auxiliary function $P(\xi )$:
\begin{equation}
P(\xi )=\langle 0| e^{\xi G} |0\rangle
\label{Pxi}
\end{equation}
then, 
\[
\langle 0| G^N |0\rangle =\left. \frac {d^N P}{d\xi ^N} \right|_{\xi =0}
\]

In order to find $P(\xi )$ we perform the following manipulations.
First, we take the derivative of $P(\xi )$:
\begin{equation}
\frac {dP}{d\xi }=\langle 0|G e^{\xi G} |0\rangle
=\langle 0| e^{\xi G} G |0\rangle
=u\: \langle 0|ab\: e^{\xi G}|0\rangle
=u\: \langle 0|e^{\xi G} a^{+}b^{+}|0\rangle
\label{deriv}
\end{equation}

Now we need to carry $a^{+}b^{+}$ over $G$ in the last expression
\begin{equation}
e^{\xi G} a^{+}b^{+}=e^{\xi G}a^{+}e^{-\xi G}e^{\xi G}b^{+}e^{-\xi G}e^{\xi G}
\label{100}
\end{equation}

It can be made by using the following equations:
\begin{eqnarray}
e^{\xi G}a^{+}e^{-\xi G} &=& a^{+}\cos \xi +(ub+ivb^{+})\sin \xi   \nonumber \\
e^{\xi G}a \: e^{-\xi G} &=& a\cos \xi  -(ub^{+}+ivb)\sin \xi   \nonumber \\
e^{\xi G}b^{+}e^{-\xi G} &=& b^{+}\cos \xi  +(ua+iva^{+})\sin \xi   \nonumber \\
e^{\xi G}b \: e^{-\xi G} &=& b\cos \xi  -(ua^{+}+iva)\sin \xi    \label{comm}
\end{eqnarray}

Substituting Eqs.(\ref{100}), (\ref{comm}) into Eq.(\ref{deriv}), we find
\[
\langle 0|e^{\xi G} a^{+}b^{+}|0\rangle=
u\sin \xi  \cos \xi \;  \langle 0|e^{\xi G}|0\rangle +
u^2 \sin ^2 \xi \;  \langle 0|ab\: e^{\xi G}|0\rangle
\]

The last equation can be rewritten as the differential equation on 
the $P(\xi )$
\begin{equation}
\frac {dP}{d\xi }=u^2 \sin \xi  \cos \xi \; P(\xi)
+u^2 \sin ^2 \xi \;  \frac {dP}{d\xi }  \label{diff}
\end{equation}
with boundary condition $P(0)=1$. 

The solution of Eq.(\ref{diff}) is
\begin{equation}
P(\xi)=\frac{1}{\sqrt{1-u^2 \sin^2 \xi }}
\end{equation}

Integrating Eq.(\ref{snorma}) over $\gamma $, we obtain
\begin{eqnarray}
\langle \Psi _0| \Psi _0\rangle =\left. \frac 12 \int_0^\pi \sin \gamma d\gamma \;
\frac {d^N P}{d\xi ^N} \right|_{\xi =0}
=\left. \frac {d^N}{d\xi ^N} \left( \frac {1}{\cos^2(\frac{\xi }{2})} 
\right)  \right|_{\xi =0}
\label{snorma1}
\end{eqnarray}

So, finally we arrive at
\begin{equation}
\langle \Psi _0| \Psi _0\rangle =
\left. 2 \frac {d^{N+1}}{d\xi ^{N+1}} \left( \tan \frac{\xi}{2}  \right)  
\right|_{\xi=0}= \frac{4\,(2^{N+2}-1)}{N+2}|B_{N+2}|
\label{snorm}
\end{equation}
Here $B_N$ are the Bernoulli numbers.

To calculate the spin correlators we need to introduce operators:
\begin{eqnarray}
G_z &=& g_{i}^{+} \, e^{izS_i^{-}} e^{iz^{\prime }S_i^{+}}2S_i^z g_{i}
= u(a^{+}b^{+}-ab)+iv(ab^{+}-a^{+}b) \nonumber \\
G_{+} &=& g_{i}^{+} \, e^{izS_i^{-}} e^{iz^{\prime }S_i^{+}}S_i^{+} g_{i}
= ua^{+}b+ivab \nonumber \\
G_{-} &=& g_{i}^{+} \, e^{izS_i^{-}} e^{iz^{\prime }S_i^{+}}S_i^{-} g_{i}
= uab^{+}+iva^{+}b^{+} \nonumber
\end{eqnarray}

Then, the correlator $\langle {\bf S}_1 {\bf S}_{l+1}\rangle$ 
will be defined by
\begin{equation}
\langle \Psi _0 |{\bf S}_1 {\bf S}_{l+1}|\Psi _0\rangle 
=\frac 12 \int_0^\pi \sin \gamma d\gamma \,
\langle 0|\: \frac 14 G_z G^l G_z G^{N-l-2}+
\frac 12 G_{+} G^l G_{-} G^{N-l-2}  |0\rangle   
\label{scor}
\end{equation}
(since $\langle 0|G_{-}...|0\rangle =0 $).

The expectation values in Eq.(\ref{scor}) can be represented as
\begin{eqnarray}
\langle 0|\: G_z G^l G_z G^{N-l-2} |0\rangle &=& 
\frac {\partial ^l}{\partial \xi ^l} \frac {\partial ^{N-l-2}}
{\partial \zeta ^{N-l-2}} 
\left. \langle 0|G_z e^{\xi G} G_z e^{\zeta G} |0\rangle
\right|_{\xi =\zeta =0}  \nonumber \\
\langle 0|\: G_{+} G^l G_{-} G^{N-l-2} |0\rangle &=& 
\frac {\partial ^l}{\partial \xi ^l} \frac {\partial ^{N-l-2}}
{\partial \zeta ^{N-l-2}} 
\left. \langle 0|G_{+} e^{\xi G} G_{-} e^{\zeta G} |0\rangle
\right|_{\xi =\zeta =0}
\label{27}
\end{eqnarray}

After the procedure similar to that for the norm and the integration 
over $\gamma $, we obtain
\begin{equation}
\langle \Psi _0 |{\bf S}_1 {\bf S}_{l+1}| \Psi _0\rangle 
= \frac {\partial ^l}{\partial \xi ^l} \frac {\partial ^{N-l-2}}
{\partial \zeta ^{N-l-2}} \left. \left( -\frac 38 \frac 
{\cos (\xi -\zeta )}{\cos^4(\frac{\xi +\zeta }{2})} \right)  
\right|_{\xi =\zeta =0}
\label{spincorr}
\end{equation}


\begin{thebibliography}{99}

\bibitem{fannes} M.Fannes, B.Nachtergaele, R.F.Werner, Commun. Math. Phys.
{\bf 144}, 443 (1992).

\bibitem{Z} A.~Klumper, A.~Schadschneider, J.~Zittartz, Z. Phys. B.
{\bf 87}, 281 (1992); Europhys. Lett. {\bf 24}(4), 293 (1993).

\bibitem{mik}  A.K.Kolezhuk, H.-J.Mikeska, Phys. Rev. Lett. {\bf 80}, 2709
(1998); Int. J. Mod. Phys. B{\bf 12}, 2325 (1998).

\bibitem{AKLT} I.~Affleck, T.~Kennedy, E.~H. Lieb, and H.~Tasaki, Phys. Rev.
Lett. {\bf 59}, 799 (1987); Commun. Math. Phys. {\bf 115}, 477 (1988).

\bibitem{JETP}  D.V.Dmitriev, V.Ya.Krivnov, A.A.Ovchinnikov, JETP {\bf 88}, 
138 (1999).

\bibitem{strack} R.Strack, D.Vollhardt, Phys. Rev. Lett. {\bf 70}, 2637 (1993),
Phys. Rev. Lett.{\bf 72}, 3425 (1994).

\bibitem{AAO}  A.A.Ovchinnikov, Mod. Phys. Lett B{\bf 7}, 1397 (1993);
J. Phys. {\bf CM 6}, 11057 (1994).

\bibitem{Boer} Jan de Boer, A.Schadschneider, Phys. Rev. Lett. {\bf 75}, 4298 
(1995).

\bibitem{Japan}  T.Hamada, J.Kane, S.Nakagawa and Y.Natsume. J. Phys. Soc. Jpn.
{\bf 57}, 1891 (1988); {\bf 58}, 3869 (1989).

\bibitem{pr}  D.V.Dmitriev, V.Ya.Krivnov, A.A.Ovchinnikov, Z. Phys. B{\bf 103}
193, (1997); Phys. Rev. B{\bf 56}, 5985 (1997).

\bibitem{KO}  V.Ya.Krivnov, A.A.Ovchinnikov, Phys. Rev. B{\bf 53}, 6435 (1996).

\bibitem{EPJ}  D.V.Dmitriev, V.Ya.Krivnov, A.A.Ovchinnikov (to be published).

\bibitem{P0}  P.van Leuven, Physica {\bf 45}, 86 (1969).

\bibitem{yang}  C.N.Yang, Rev. Mod. Phys. {\bf 34}, 694 (1962).

\bibitem{yang1}  C.N.Yang, Phys. Rev. Lett. {\bf 63}, 2144 (1989).

\bibitem{eta}  F.H.L.Essler, V.E.Korepin, K.Schoutens, Phys. Rev. Lett. 
{\bf 68}, 2960 (1992);  Phys. Rev. Lett. {\bf 70}, 73 (1993). 

\end{thebibliography}
\end{document}